# *On Origins of Alpha*

Zura Kakushadze[§†1]

[§] *Quantigic® Solutions LLC,[2] 1127 High Ridge Road, #135, Stamford, CT 06905*

[†] *Free University of Tbilisi, Business School & School of Physics*
*240, David Agmashenebeli Alley, Tbilisi, 0159, Georgia*

June 16, 2015

## Abstract

We argue that an important contributing factor into market inefficiency is the lack of a robust mechanism for the stock price to rise if a company has good earnings, *e.g.*, via buybacks/dividends. Instead, the stock price is prone to volatility due to rather random perception/interpretation of earnings announcements (among other data) by market participants. We present empirical evidence indicating that dividend paying stocks on average are less volatile, even factoring out market cap. We further ponder possible ways of increasing market efficiency via 1) instituting such a mechanism, 2) a taxation scheme that would depend on holding periods, and 3) a universal crossing engine/exchange for mutual and pension funds (and similar long holding horizon vehicles) with no dark pools, 100% transparency, and no advantage for timing orders.

---

According to the SEC (U.S. Securities and Exchange Commission) website,[3] "A Ponzi scheme is an investment fraud that involves the payment of purported returns to existing investors from funds contributed by new investors." SEC further clarifies: "With little or no legitimate earnings, Ponzi schemes require a consistent flow of money from new investors to continue. Ponzi schemes tend to collapse when it becomes difficult to recruit new investors or when a large number of investors ask to cash out." So, is the stock market a Ponzi scheme? (See Gross [2012].) As we will argue below, the truth appears to have various shades of grey…

Ultimately, our goal here is to ponder the origins of alpha. Why does alpha exist? Where does it come from? The stock market is man-made, with all its complex, artificial and not always rational rules and regulations. Unlike, say, the planetary motion in our solar system, which is governed by the fundamental laws of physics (to wit, gravity), the stock market is messier. It is not directly governed by the fundamental laws of nature.[4] This lack of underlying fundamental description makes the origins of alpha murkier. But not all is lost. Let us start by asking:

**Why does anyone invest in stocks?**

"Greed!" some would proclaim. More moderately, the idea is simple. You buy a stock and hope its price goes up, so at some point you can sell it at a profit. But what is this "hope" based on? The stock[5] price – at least in free market – is determined by supply and demand. If the company that issued the stock has substantial earnings and accumulates extra cash on its books, then it can buy back its own stock and drive the price up. However, there is no law, rule or regulation that would require any company to buy back its own stock or pay dividends[6] – another way of rewarding its shareholders – if it has good earnings or excess cash. The decision to buy back the stock or issue dividend is in the sole discretion of the company's board of directors and does not require the shareholders' approval. On paper, shareholders have voting power and can – in theory – replace the board of directors if they are unhappy with it. However, in practice only large shareholders have such power. Small (including most individual) shareholders are effectively disenfranchised in this regard, with no guarantee of stock buybacks or dividends.

---

[3] http://www.sec.gov/answers/ponzi.htm, accessed January 7, 2015.
[4] Nor is there any fundamental law of nature that would dictate that stocks should exist in the first instance.
[5] Here we focus on common stock. The existence of preferred stock does not alter the conclusions.
[6] Typically, annualized stock dividend is a low single digit percentile of the stock price. Therefore, dividends are not the primary mechanism for profiting from stock investments. Most U.S. companies do not even pay dividends.

With no such guarantees, the aforementioned "hope" that the stock price will rise is based on the expectation that, if the company has good earnings, other investors will want to purchase its stock. However, again, with no such guarantees, naturally, one may ask:

**Why do earnings affect the stock price?**

The evident answer is that the market movers and shakers base their decisions on the earnings, and then the majority of the rest of the market participants follow the suit. And the reason why the former look at the earnings is that a subset of companies – typically these are established companies with solid track records – follow *unwritten rules* of rewarding their shareholders via buybacks and/or dividends when their earnings are good. Many of these companies have large shareholders – these are some of the aforementioned movers and shakers, including those that have been grandfathered down – so for such companies it makes perfect sense to reward shareholders with buybacks and dividends. However, many – in fact, most – companies have no intention of following the aforementioned unwritten rules. They do not have solid track records and their stock prices follow the earnings only if – and to the extent that – most market participants extrapolate basing their decision-making on earnings to such less established companies from the way they treat the established companies that actually follow the unwritten rules. So, in a way, most companies then would appear to be "free-riding".

The lack of a guaranteed mechanism for the shareholders being rewarded by the companies themselves has undesirable consequences. Thus, it makes it tempting for companies to cook their books and exaggerate their earnings, which drive the stock prices. If a company was required to buy back its own stock and/or issue dividend when its earnings are good, it would have much less incentive to exaggerate its earnings – had such a requirement existed, the dishonest accounting practices during the dotcom era, when many companies had no intention to play by the unwritten rules and instead aimed at making a quick buck at the shareholders' expense, likely would not have been as widespread. Also, since only a fraction of the companies follows the unwritten rules, bubbles are likely unavoidable as investments in the rest of the market are based on the hope that the corresponding stock prices will rise due to more money being pumped into such stocks by new investments, not via, say, buybacks. If it were not for the aforementioned established companies, it would be a (*quasi*) Ponzi scheme.

**Where does alpha come from?**

The above discussion helps – we hope – shed light on the origins of alpha. With everything else being equal, the market likely would be more efficient if all companies followed the unwritten rules – or if they were required to reward their shareholders by systematically increasing the stock price via, say, buybacks with a formulaic relation between the required buybacks and earnings (among other details). As it stands, the supply and demand is driven by what appears to be a rather random perception/interpretation of the earnings announcements (among other information) by market participants. This leads to volatility and mispricings at various time horizons. These mispricings are then arbitraged away by what can be generically termed as "mean-reversion" (or "contrarian") strategies. For a given "mean-reversion" time horizon there might also exist opportunities to profit via what can be generically termed as "momentum" strategies on accordingly shorter (and, in some cases, longer) time horizons.

      One important ingredient that is implicitly assumed in the above discussion is market impact and executions. Even if every company under the Sun followed the unwritten rules, due to a large number of market players and virtual impossibility to predict supply and demand imbalances or their precise timings, mispricings and inefficiencies are inevitable. Longer horizon strategies thereby create arbitrage opportunities on somewhat shorter horizons; strategies on such scales create arbitrage opportunities on yet shorter time scales; and so on – all the way down to HFT (high frequency trading) strategies. While on the longest horizon time scales the strategies are mostly long (mutual and pension funds, holding companies, *etc*.), on shorter horizons strategies can be dollar neutral, hence seemingly creating profit "out of thin air" – which does not take into account substantial monetary and human capital involved.

**More market efficiency, less alpha?**

Along with transactions costs, dollar neutral and, more generally, proprietary strategies[7] that operate on shorter scales make the stock market deviate from a "zero-sum" game between passive and active investments strategies – if such investments are restricted to long-only strategies, that is. Both passive and active investment strategies "feed" those who profit from transaction costs as well as proprietary strategies. Once the latter are taken into account, it is indeed a "zero-sum" game.[8] However, the money does not appear out of thin air, the profits made, say, by proprietary strategies ultimately come from the mutual and pension funds and other long-only investments, including retail traders. A mom-and-pop/grandma-and-grandpa investor advocate may wonder, what – if anything – can be done to keep that money from trickling down. It appears that there are some ways of achieving this, albeit they may not find enthusiastic support among various groups with vested interests in the subject. Here is a list:

> 1. If companies are required to do buybacks and/or issue dividends in a simple formulaic fashion when their earnings are good – *e.g.*, proportionately to their earnings above some threshold – this would likely reduce volatility and arbitrage opportunities in the market. It would also likely reduce the companies' desire to cook their books.
> 
> 2. If there is a progressive tax on trading profits – the shorter the holding period, the higher the tax – this would likely make HFT and similar proprietary trading much less appealing. Naturally, imposing such progressive taxes would be politically challenging.[9]
> 
> 3. A universal crossing engine/exchange with no dark pools, 100% transparency, and no advantage for timing orders, would likely dramatically reduce transaction costs for large mutual and pension funds and similar vehicles. Ultimately, it would also likely reduce arbitrage opportunities on shorter time horizon scales by reducing volatility.[10] Likely, this would substantially weaken the argument that proprietary strategies are a necessity for they provide liquidity, price discovery and make the market more efficient.

---

[7] Dollar (or "market") neutrality is not crucial in the context of the following discussion. What is important is that profits are made from arbitrage opportunities, not via long-horizon investments into companies.

The ideas listed above may not be all-around popular or perfect. They are not intended as such. Hopefully, they are thought provoking. After all, what is the key idea behind the stock market? Arguably, it is for companies to raise money – in some sense, it is a form of crowdfunding but with a hope of a profit – to invest into their businesses, create new products and services, make life better for everyone and grow the economy. Then the question is how to structure the stock market in a way that would make it work most efficiently, albeit imperfectly, for this noble goal.

---

[8] See Sharpe [1991].
[9] One would also have to ponder international arbitrage opportunities this could create.
[10] For such a universal crossing engine/exchange to work, it would not suffice to "slow down" incoming orders (*cf.* Lewis [2014]). The orders would have to be aggregated in a 100% transparent fashion (so all market participants can see all orders) over a long enough period such that arbitraging these orders makes as little sense as possible.

**What about empirical evidence?**

Since there is no law, rule or regulation that would require any company to buy back its own stock or pay dividends, direct empirical measurements of the effects of such a requirement on alphas or existence thereof is not an option. However, not all is lost. We can design a test to see, *e.g.*, if there is a correlation between (not) paying dividends and the "extent" of alpha (arbitrage) opportunities a stock creates. However, first we must quantify what this "extent" entails.

In this regard, it is important to note that we are not interested in the stock performance (or drift in the stochastic language). Instead, we are interested in the stock *volatility*, *i.e.*, we wish to see whether there is a correlation between (not) paying dividends and the stock volatility. Here some care is needed as the logarithmic stock volatility is strongly anti-correlated with the logarithmic market cap, and there is a strong correlation between paying dividends and logarithmic market cap as well. Therefore, we run a cross-sectional regression of the logarithmic historical stock return volatility over a 2-column matrix (plus the intercept). The first column is binary: its element equals 1 if the corresponding stock pays dividends pursuant to the definition below; otherwise, it is 0. The second column is populated by log of the market cap for each stock. The stock returns, for which the historical volatility is computed, are the overnight (previous close to open) returns – we have taken such a short time horizon deliberately, to wit, to ensure that the time horizons for the returns (overnight) and the dividends (quarterly) are as uncorrelated as possible. We take the daily overnight returns for the 4-year period 9/2/2010-9/4/2014, inclusive (1008 trading days, to be precise). We take market cap as of 9/1/2010.[11] The stock is considered to pay dividends if it has paid at least 16 nonzero dividends during the period 6/1/2010-12/1/2014, *i.e.*, we "pad" the aforementioned 4-year period by about a quarter on each end (to avoid missing the dividends that were paid just before or right after the 4-year period) and require that the number of paid dividends be one per quarter. The universe of stocks we take for our test consists of the U.S. listed common stocks and class shares (no OTCs, preferred shares, *etc.*) which have pricing data on

---

[11] Taking market cap out-of-sample (as we have done) or (partially) in-sample has only a small effect on the results.

http://finance.yahoo.com as of (and accessed on) 9/6/2014.[12] The dividend and market cap data is also available from the same source. For comparison purposes we also run the same regression by requiring 1 to 10 nonzero dividends in the above definition. The results are given in Table 1.

| Coefficients | Estimate | Standard Error | t-value:16+ dividends | t-value:1-10 dividends |
|---|---|---|---|---|
| Intercept | -3.753 | 0.1245 | -30.16 | -21.54 |
| Dividends (binary) | -0.714 | 0.0268 | -26.66 | 3.728 |
| Market Cap | -0.232 | 0.0063 | -36.67 | -45.26 |

**Table 1.** The regression coefficients (estimates), standard errors and t-values for the regression of the logarithmic historical volatility over the binary dividends (1 if at least 16 dividends are paid, and 0 otherwise – see above) and log of the market cap (plus the intercept). The total number of stocks is 3422. The number of stocks with 16+ (1-10) dividends is 1266 (588); the F-statistic is 2306 (1043); the multiple and adjusted R-squared are 0.4027 (0.3789) and 0.4025 (0.3786).

Our test results indicate a strong negative correlation between log of the stock volatility and whether the stock pays dividends, *i.e.*, dividend paying stocks on average are less volatile. A similar analysis for buybacks requires buyback announcements (both dates and sizes),[13] which is not freely available data, and it is outside of the scope of this note. We hope to perform such analysis in the future.

**References**

Fruin, P. and Ma, L. "Buying Outperformance: Do Share Repurchase Announcements Lead to Higher Returns?" S&P Capital IQ, McGraw Hill Financial, Quantamental Research Report,

---

[12] The choice of the 4-year window is based on what data was readily available.
[13] There is a positive correlation between buybacks and stock performance; see Fruin and Ma [2014] and references therein.

January 23, 2014. Available online from: http://www.spcapitaliq.com/our-thinking/research.html?daterange=2014.

Gross, W.H. "Cult Figures." PIMCO Investment Outlook, August, 2012. Available online from: http://www.pimco.com/EN/Insights/Pages/Cult-Figures.aspx.

Lewis, M. "Flash Boys: A Wall Street Revolt." New York, NY: W.W. Norton & Company, Inc., 2014.

Sharpe, W. "The Arithmetic of Active Management." *Financial Analysts' Journal*, 47 (1991), pp. 7-9.